\documentclass[12pt]{iopart}

\usepackage{graphicx}
\begin{document}

\title[Motion of run-and-tumble particles]{Non-Gaussian anomalous dynamics in systems of interacting run-and-tumble particles}

\author{Stefanie Put$^1$, Jonas Berx$^2$ and Carlo Vanderzande$^{1,2,\dagger}$}

\address{$^1$ Faculty of Sciences, Hasselt University, Agoralaan 1, 3590 Diepenbeek, Belgium \\
$^2$ Instituut Theoretische Fysica, KULeuven, Celestijnenlaan 200D, 3001 Heverlee, Belgium \\
$\dagger$ Deceased September 2, 2019}
\ead{stefanie.put@uhasselt.be}
\vspace{10pt}

\begin{abstract}
The motion of a tagged degree of freedom can give important insight in the interactions present in a complex environment. We investigate the dynamics of a tagged particle in two non-equilibrium systems that consist of interacting run-and-tumble particles. The first one is an exactly solvable polymer model, the second is  a two-dimensional lattice model, which is studied through simulations. We find that in both cases a tagged particle shows anomalous dynamics and non-Gaussian behaviour for times below the persistence time of the run-and-tumble motion. For later times, the dynamics of the tagged monomer becomes diffusive and Gaussian. In the lattice model, non-Gaussianity persists and can, for intermediate densities, be well approximated by a Laplace distribution. We attribute this behaviour to the dynamically changing environment of the tagged particle, which we argue, is an essential ingredient to observe deviations from Gaussianity.
\end{abstract}

%
%
%
%
%

\section{Introduction}
The study of the dynamics of a tagged particle in a fluid goes back to the experiments of Brown on the random motion of pollen grains in water \cite{Brown28} and to the explanation of these experiments in terms of collisions with particles of the embedding fluid by Einstein \cite{Einstein05}, Langevin \cite{Langevin08} and others. From this work, it became clear that following the motion of a tagged particle such as a colloid or small polymer in a simple liquid or gas allows one to obtain insight in the interactions that are present in the fluid. This is the idea behind passive rheology. 
Moreover, using linear response theory and projection techniques \cite{Zwanzig61,Mori65}, it is possible to derive the more heuristic theories of Brownian motion from a fundamental statistical mechanical perspective. Using this approach it becomes possible to extend the theory of Brownian motion to other slow coordinates such as, for example, the reaction coordinates used to describe the progress of a chemical reaction or the folding of a biopolymer \cite{Zwanzig01}. 

Turning back to Brownian motion, it is well established that in a simple fluid, the position of a colloidal particle at time $t$ is a random variable with a Gaussian distribution whose variance scales linearly with time. Modern microscopic techniques have allowed to extend studies of Brownian motion to more complex environments such as living cells \cite{Golding08,Weigel11,Jeon11}, membranes \cite{Kneller11,Jeon12} and artificially crowded fluids \cite{Szymanski09,Jeon13}. It has, for example, been found that crowdedness leads to viscoelastic behaviour in which friction forces have a long-time memory \cite{Weiss13}. This in turn gives rise to motion that is subdiffusive, i.e. the variance of the position of a tagged particle grows as $t^\beta$ where $\beta<1$. Besides crowdedness, heterogeneity is another property of complex (biological) environments. It has been argued theoretically and observed experimentally that this heterogeneity can lead to displacement distributions that are non-Gaussian \cite{Metzler14, Wang12,Wang09,Toyota11,Parry14,Chechkin17}. 

Most of the studies mentioned so far concerned systems that are in thermal equilibrium. Much less is known about the motion of a Brownian particle, a reaction coordinate or other slow variables in systems that are out of equilibrium. One reason for this is that there is at present no complete theory of non-equilibrium statistical mechanics and therefore the necessary theoretical framework of linear response, projection techniques, ... is much less developed \cite{Maes14,Maes15,Steffenoni16}. However, recent years have seen considerable progress in understanding some of the most simple types of non-equilibrium systems like for example low-dimensional exclusion processes where one or more driven particles move through a background of passive diffusing particles. These systems show a rich phenomenology including enhanced diffusivity and negative differential mobility \cite{Benichou14,Basu14,Baiesi15,Illien18}. Also superdiffusive behaviour (for which $\beta> 1$) has been observed \cite{Benichou13}. 

Another field in out-of-equilibrium condensed matter research where much progress has been achieved in the past years is that of active matter. An active particle is a particle that consumes free energy (such as that made available by the hydrolysis of ATP) to generate motion \cite{Marchetti13,Bechinger16}.  Several bacterial species such as {\it E. Coli} perform a so called run-and-tumble motion in which they move with a constant velocity in a fixed direction \cite{Berg04}. With a rate $\lambda$ the direction of this motion changes randomly. A simple mathematical description is in terms of a persistent random walk \cite{Schnitzer93}. 

In the present Letter we want to investigate in detail the motion of a tagged particle in two models that contain run-and-tumble particles in the absence or presence of a thermal bath. The first one is a simple bead-spring polymer model whose monomers perform run-and-tumble motion and are subject to thermal noise. While this is not a realistic model for any (bio)polymer it has the advantage that it is an interacting particle model that to a large extend can be studied analytically. This allows us to get exact results for the mean squared displacements of the monomers. In the second model, run-and-tumble particles hop on a lattice and the jump rates contain a thermal and a non-equilibrium component.

Recently the motion of a run-and-tumble particle in one dimension has received considerable interest \cite{Malakar18}. Special attention was given to the probability distribution of the particle position. Our work can also be seen as an extension of that research to systems with many interacting run-and-tumble particles.

This paper is organised as follows. In section 2 we present our polymer model with run-and-tumble monomers. We calculate exactly the mean squared displacement of a monomer and find that it can display ballistic, superdiffusive, subdiffusive and diffusive regimes depending on temperature and time. We also calculate the kurtosis and the whole probability distribution of the displacement in the various regimes from a simulation of the model. Clear evidence for non-Gaussian behaviour on time scales below the persistence time $\tau=1/\lambda$ shows up. In section 3 we introduce our lattice model of run-and-tumble particles. At zero temperature it coincides with the persistent exclusion process (PEP) introduced in \cite{Soto14}. From extensive simulations we derive the mean squared displacement, the kurtosis and the whole probability distribution for the position of a tagged active particle. In this model we find non-Gaussian behaviour for the whole time regime investigated. We trace this behaviour to a dynamically changing environment for the tagged particle. In section 4 we present our conclusions. 
\section{Rouse model with run-and-tumble monomers}
As already mentioned in the introduction a run-and-tumble particle moves (or {\it runs}) with constant velocity $v$ in a fixed direction given by a unit vector $\hat{e}(t) $. At a rate $\lambda$ this direction changes (or {\it tumbles}) into another one chosen uniformly on the sphere. In addition, the particle may be subject to a thermal noise $\vec{\eta}(t)$. In an overdamped regime the motion of the particle is given by the Langevin equation
\begin{eqnarray}
\gamma \frac{d\vec{x}}{dt} = v \hat{e}(t) + \vec{\eta}(t),
\end{eqnarray}
where $\gamma$ is the friction coefficient. We assume that the thermal noise is a Gaussian stochastic process with average $\langle\vec{\eta}(t)\rangle=0$ and correlation given by the fluctuation-dissipation theorem
\begin{eqnarray} \label{eq:thermal}
\langle \vec{\eta}(t)\cdot \vec{\eta}(t') \rangle=6 \gamma k_B T \delta(t-t').
\end{eqnarray}
For the tumbling process we have $\langle \hat{e}(t)\rangle=0$ and $\langle \hat{e}(t)\cdot\hat{e}(t')\rangle=e^{-\lambda|t-t'|}$. This process is non-Gaussian. 

We now make a polymer of $N$ such run-and-tumble particles by connecting them with springs of spring constant $k$. We denote the position of the $n$th monomer by $\vec{R}_n$ ($n \in \{0,1,\ldots,N-1\}$).  Its equation of motion is
\begin{eqnarray} \label{eq:Langevin}
\gamma \frac{d\vec{R}_n}{dt} =-k \left[2 \vec{R}_n(t) - \vec{R}_{n-1}(t) - \vec{R}_{n+1}(t) \right] + v \hat{e}_n(t) + \vec{\eta}_n(t).
\label{3}
\end{eqnarray}
The directions of motion $\hat{e}_n(t)$ and the thermal noises $\vec{\eta}_n(t)$ of different monomers are independent random variables. The Langevin equations (\ref{3}) are supplemented by the boundary conditions $\vec{R}_{-1}(t)=\vec{R}_0(t)$ and $\vec{R}_{N-1}(t)=\vec{R}_N(t)$. For $v=0$ this model reduces to the Rouse model \cite{Rouse53}, which is the simplest model in polymer dynamics since it neglects excluded volume and hydrodynamic interactions \cite{Doi86}. Our Rouse model with run-and-tumble monomers has no immediate relevance to polymer physics but can be studied as an exactly solvable model of interacting active particles. Recently similar models of polymers subject to active forces have been studied \cite{Vandebroek15,Sakaue16,Samanta16,Osmanovic17,Osmanovic18}. These studies however mostly considered the case in which the active forces are given by (Gaussian) Ornstein-Uhlenbeck processes. This can also be interpreted as a Rouse model in which the monomers are active Brownian particles. The latter form a class of active particles distinct from run-and-tumble particles in which the direction of the velocity changes because of rotational diffusion.

\subsection{Super- and subdiffusive motion of the monomers}
The set of equations (\ref{3}) with the given boundary conditions can be solved by introducing the normal, or Rouse, modes $\vec{X}_p(t)$ ($p \in \{0,\ldots,N-1\}$)
\begin{eqnarray}
\vec{X}_p(t) = \frac{1}{N} \sum_{n=0}^{N-1} c_n^p \vec{R}_n (t),
\label{4}
\end{eqnarray}
where $c_n^p=\cos\left[\pi p(n+1/2)/N\right]$. 
Using standard techniques (see \ref{appendix:normModes}) one can determine the time evolution of these normal modes in terms of the thermal and active forces. From these one finds the correlations between the modes. The result is
\begin{eqnarray}
\langle \vec{X}_p(t)\cdot \vec{X}_q(t') \rangle = \left[\frac{3 k_B T}{k_p} e^{-|t-t'|/\tau_p}  + \frac{v^2}{2 N \gamma^2} C_p(t,t')\right] \delta_{p,q}.
\label{5}
\end{eqnarray}
The first term is the thermal contribution. Here $k_p=8 N k \sin^2 \left(\pi p/2 N\right)$ while the relaxation time $\tau_p$ of the $p$th mode equals $2 N \gamma/k_p$. The longest relaxation time $\tau_1$ will be referred to as the Rouse time $\tau_R\equiv\tau_1 \approx \gamma N^2/\pi^2 k$ ($N \gg 1$). It corresponds physically to the time it takes for the polymer to diffuse over its own radius of gyration due to thermal forces only. The second term is the contribution of the active motion for which the time dependent function $C_p(t,t')$ is given by
\begin{eqnarray}
C_p(t,t') =\int_0^t \int_0^{t'} e^{-\lambda |t-\tau-t'+\tau'|} e^{-(\tau+\tau')/\tau_p}\ d\tau\ d\tau',
\end{eqnarray}
an integral that can easily be determined but whose lengthy expression is not given here. 

From the inverse of (\ref{4}),
\begin{eqnarray}\label{eq:naturalcoords}
\vec{R}_n(t) = \vec{X}_0(t) + 2 \sum_{p=1}^{N-1} c_n^p \vec{X}_p(t),
\end{eqnarray}
and by using the correlation (\ref{5}), one can determine (see (\ref{appendix:msd})) the mean squared displacement of the $n$th monomer, $\Delta R_n^2(t) \equiv \langle \left[\vec{R}_n(t) - \vec{R}_n(0) \right]^2\rangle$. The result is
\begin{eqnarray}
\Delta R_n^2(t)  &=& \frac{6k_B T}{N \gamma}\left[t + 2 \sum_{p=1}^{N-1} \left(c_n^p\right)^2 \tau_p \left(1-e^{-t/\tau_p}\right)\right] \nonumber \\
&+& \frac{v^2}{N \gamma^2}\left[I_0(t) + 2 \sum_{p=1}^{N-1} \left(c_n^p\right)^2  C_p(t,t)\right].
\label{8}
\end{eqnarray}
Here 
\begin{eqnarray}
I_0(t)=\int_0^t \int_0^{t} e^{-\lambda |\tau-\tau'|} d\tau\ d\tau'=\frac{2(-1+\lambda t + e^{-\lambda t})}{\lambda^2}.
\end{eqnarray}
Clearly in (\ref{8}), the first line is the thermal contribution, the second the active one. For each of these, the first term is the contribution of the centre of mass of the polymer, the second that of the individual monomers. 

The motion of a monomer in a Rouse chain with passive monomers ($v=0$) is well established \cite{Doi86,Briels94}. It is diffusive both in the initial time regime and asymptotically in time. This can be easily seen by respectively taking the lowest order Taylor series in $t$ and the long time limit of the first line of (\ref{8}). Physically, for small times a monomer doesn't notice the presence of the other monomers, while at late times each monomer has to follow the diffusive motion of the centre of mass. For intermediate times $\tau_N < t < \tau_R$ one can show that $\Delta R_n^2(t)  \sim t^{1/2}$ by approximating the sum over $p$ by an integral and performing a change of variables. In this time window the passive monomer therefore subdiffuses. 

For a polymer consisting of active monomers at $T=0$ a similar analysis can be made. The run-and-tumble motion introduces an new time scale $\tau=\lambda^{-1}$. For long chains it is reasonable to assume that $\tau_N < \tau < \tau_R$. From a lowest order Taylor expansion of the second line of (\ref{8}) one sees that initially $\Delta R_n^2(t)  \sim t^2$ which is the ballistic motion of a run-and-tumble particle. This motion lasts untill $t \approx \tau_N$ after which the interaction with other monomers becomes relevant.  It is possible to show (see (\ref{appendix:superdiff})) that in the next time regime, $\tau_N < t < \tau$,  the mean squared displacement shows superdiffusive behaviour with an exponent $3/2$. 

On timescales much larger than $\tau$ we can approximate the colored active noise as a white noise. We therefore expect that the behaviour becomes similar to the thermal one, i.e. subdiffusive monomer motion for $\tau \ll t<\tau_R$ and diffusive motion for $t> \tau_R$. In Figure 1, we have plotted the exact result (\ref{8}) at $T=0$ for the middle monomer in a chain of $N=513$ monomers. One clearly sees the four time regimes: ballistic ($t<\tau_N$), superdiffusive ($\tau_N < t <\tau$), subdiffusive ($\tau \ll t < \tau_R$) and diffusive $t> \tau_R$.
The deviation of the exponents from the expected ones is caused by finite size effects. From the above reasonings it is also clear that for $\tau > \tau_R$, the subdiffusive regime disappears. 

\begin{figure}[!h]
\centering
  \includegraphics[width=7cm]{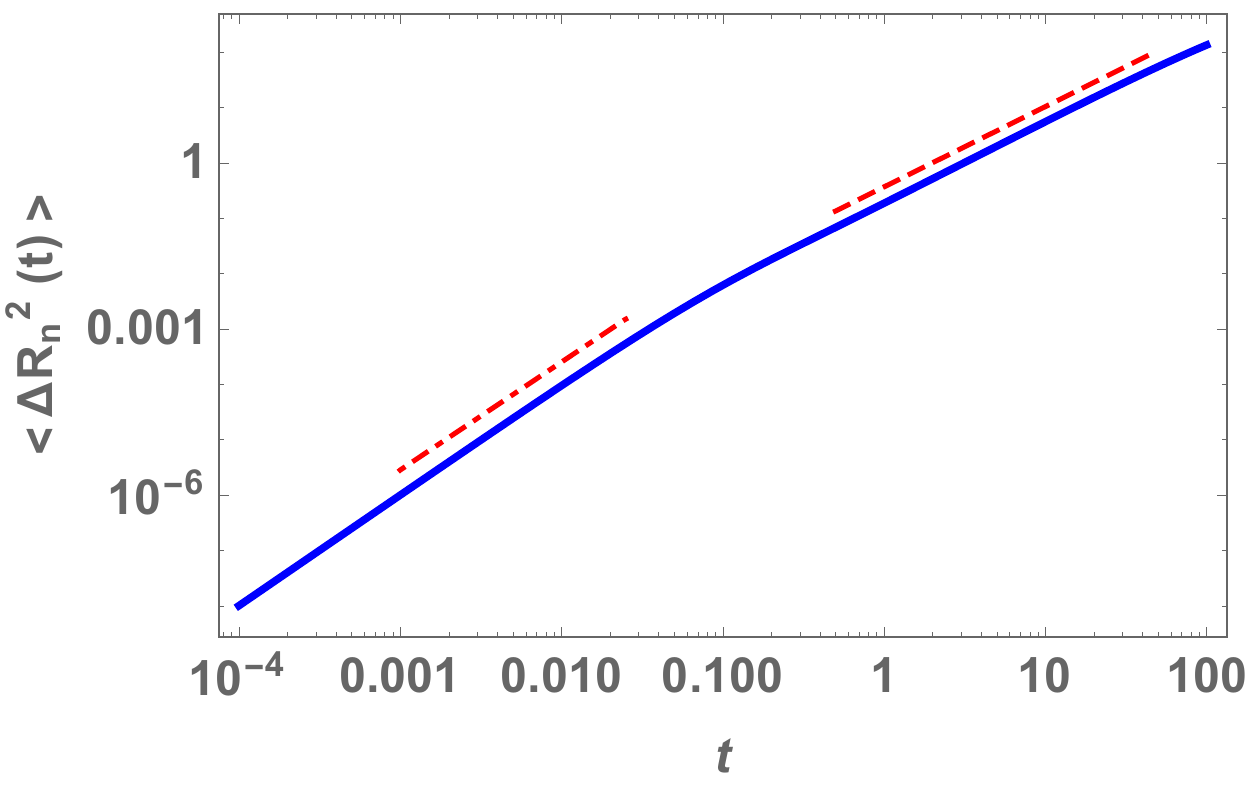}
  \includegraphics[width=7cm]{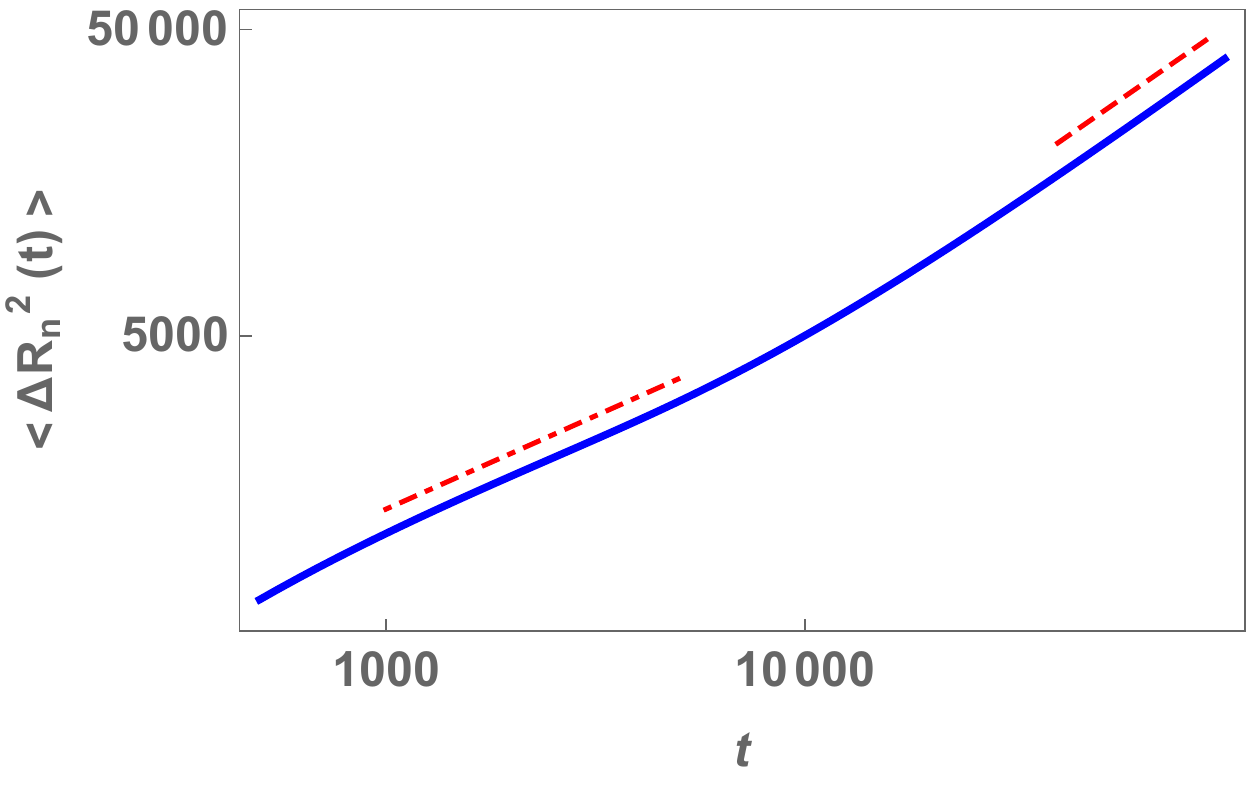}
\caption{Mean squared displacement $\Delta R_n^2(t)$ of the middle monomer ($n=256$) as a function of time in a chain with $N=513$, $T=0, v=\gamma=1$. For the parameters used in this plot we have $\tau_N=0.08, \tau=100, \tau_R \approx 9000$. On the left side, we show the early time regime $t< \tau$. The dot-dashed and dashed lines have slope $1.95$ and $1.45$ corresponding with the ballistic and superdiffusive regime. On the right side, we plot the regime $t>\tau$. The dot-dashed and dashed lines have slope $0.61$ and $0.96$ corresponding to a subdiffusive and diffusive regime.}
\end{figure}

Next, we look at the motion of the tagged monomer when both thermal and active processes are present. Since for $t > \tau$ both behave similarly their effect just adds and we keep both the subdiffusive and the diffusive regime but with an enhanced amplitude compared to the $T=0$ case. At sufficiently small times, the thermal diffusion always dominates over the active ballistic motion since for $t$ small one has $t>t^2$. The behaviour at later times depends on the temperature and on the persistence time of the run-and-tumble motion. For temperatures that are not too high and for $\tau>1$ we expect a crossover to a superdiffusive regime for $1<t<\tau$. In Fig. 2 we show, for $\tau=200$, the mean squared displacement as a function of time for $T=0, 0.25, 1$ and $2$. We see the expected behaviour: as soon as $T \neq 0$, the early time motion is diffusive. After a subdiffusive regime we recover the superdiffusive one, albeit with an effective exponent that is smaller than $3/2$ due to the thermal effects. The width of the window in which the superdiffusion occurs decreases with increasing temperature and disappears at high temperatures. In conclusion, at temperatures that are not too high we have five regimes: diffusive, subdiffusive, superdiffusive, subdiffusive and diffusive. At high temperatures we recover the purely thermal behaviour with two diffusive regimes that are separated by a subdiffusive one. 

\begin{figure}[!ht]
\centering
\includegraphics[width=12cm]{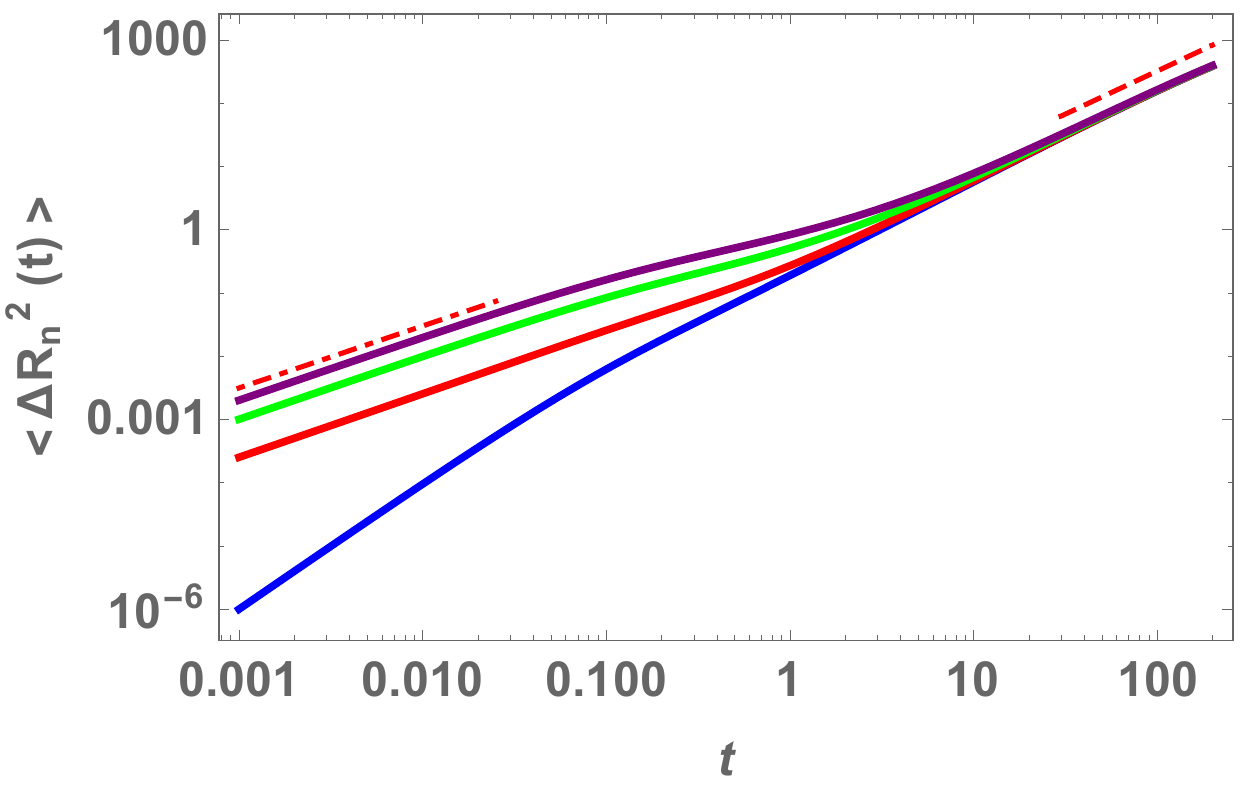}
\caption{Mean squared displacement of the middle monomer as a function of time in a chain with $N=513$. The graphs are for $T=0, 0.25, 1$ and $2$ (bottom to top), $v=\gamma=6k_B=1$. The dot-dashed line has a slope $0.98$ (diffuse regime) while the dashed line has slope $1.32$ (superdiffusive regime). For the parameters used in this plot we have $\tau_N=.08, \tau=200, \tau_R \approx 9000$.}
\label{fig2}
\end{figure}

\subsection{Kurtosis and non-Gaussianity}
We want to further characterise the various regimes that we encountered and study whether they are Gaussian or not. For this purpose we determined the kurtosis of the monomer displacement and its whole distribution from a numerical integration of the Langevin equations (\ref{3}). 

For a random variable $X$, the kurtosis $\kappa$ is defined as $\kappa=\langle(X-\mu)^4\rangle/(\langle(X-\mu)^2\rangle)^2$ where $\mu=\langle X \rangle$. For the Gaussian distribution, it equals $3$. Deviations from this value can therefore be used as an indication of non-Gaussianity.
Here we study the kurtosis for the displacement in the $x$-direction of the middle monomer in the chain. Hence we take $X\equiv\Delta \vec{R}_{x,N/2}(t)=\vec{R}_{x,N/2}(t) - \vec{R}_{x,N/2}(0)$. 

\begin{figure}
\centering
\includegraphics[width=12cm]{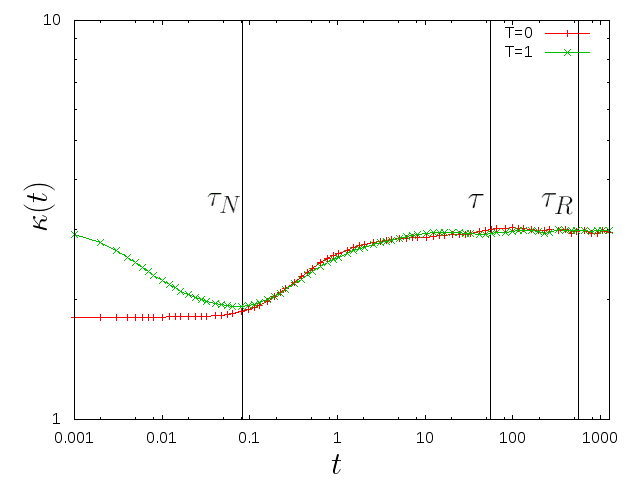}
\caption{Kurtosis $\kappa(t)$ in a chain with $N=128$ as obtained from a numerical integration of the Langevin equation (\ref{3}). The green (x) line represents our data for $T=1$, while the red (+) line are those at $T=0$. The simulation uses dimensionless units: $\gamma = 1$, $v = 50$ and $k=3$. For this simulation $\tau_N = 1/12$ and $\tau_R = 553.3$. We chose $\tau= \tau_R/10 = 55.3$ and averaged over 50000 histories.}
\label{fig:kurtosis}
\end{figure}

We present our results for $\kappa(t)$ in Fig. \ref{fig:kurtosis}. At finite temperature ($T=1$, green line) we first thermalise the chain in the absence of active forces. When those are switched on at $t=0$, we see that $\kappa$ starts deviating from $3$, a clear indication of non-Gaussianity. The kurtosis reaches a minimum around time $t \approx \tau_N$.  For later times, $\kappa(t)$ increases again and approaches the Gaussian value around $t \approx \tau$. For $T=0$ (red line), the chain cannot be equilibrated and the motion is determined by active forces only. This causes a difference in the early time behaviour but after $t \approx \tau_N$ we notice little difference between the situation at zero temperature and that at finite temperature.

\begin{figure}
\centering
  \includegraphics[width=7cm]{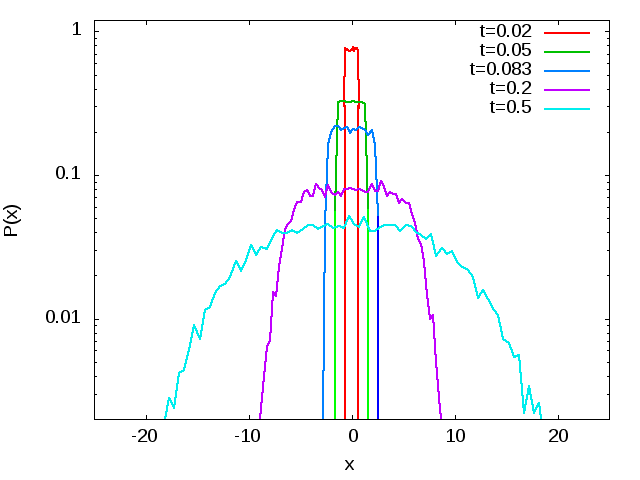}
  \includegraphics[width=7cm]{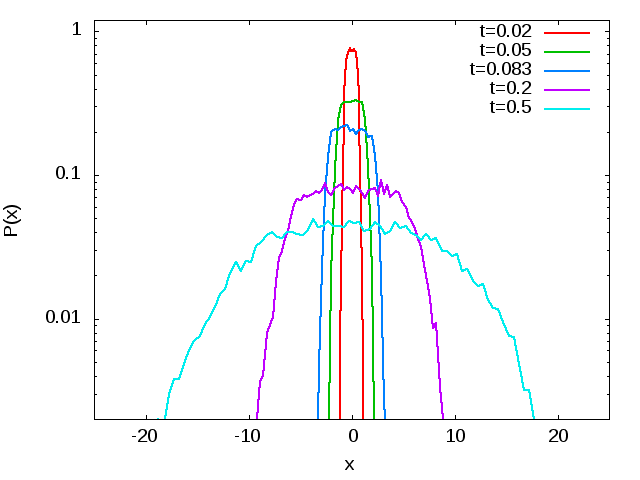}
\caption{The probability density $P(X)$ (with $X=\Delta \vec{R}_{x,N/2}(t))$) at different times in a chain with $N=128$. The distributions are shown for times $t \ll \tau$. The simulation uses dimensionless units: $\gamma = 1$, $v = 50$, $k=3$. For these parameter values $\tau_N=1/12 \approx 0.083$. We chose $\tau= \tau_R/10 = 55.3$ and averaged over 50000 histories. Left side: results for $T=0$. Right side: results for $T=1$.}
\label{fig:korteTijden}
\end{figure}

\begin{figure}
\centering
  \includegraphics[width=7cm]{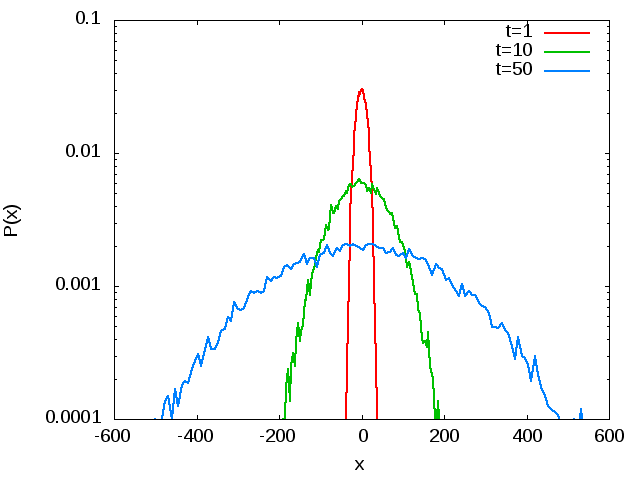}
  \includegraphics[width=7cm]{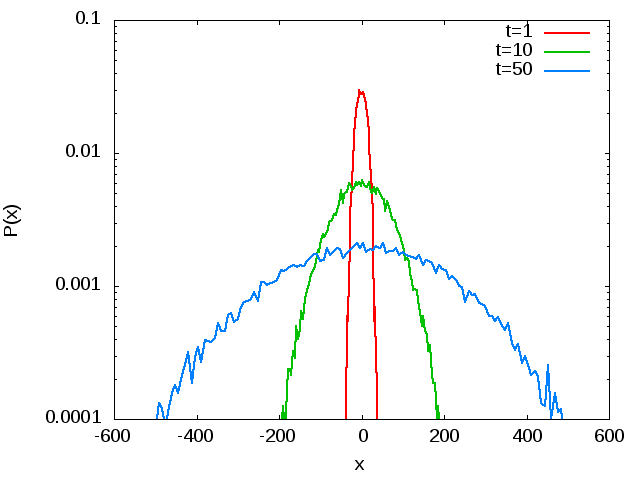}
\caption{The probability density $P(X)$ (with $X=\Delta \vec{R}_{x,N/2}(t))$) at different times in a chain with $N=128$. The distributions are shown for times $\tau_N \ll t < \tau$. The simulation uses dimensionless units: $\gamma = 1$, $v = 50$, $k=3$. For these parameter values $\tau_N=1/12 \approx 0.083$. We chose $\tau= \tau_R/10 = 55.3$ and averaged over 50000 histories. Left side: results for $T=0$. Right side: results for $T=1$.}
\label{fig:langeTijden}
\end{figure}

We can find more evidence for this non-Gaussian behaviour in the probability density distribution of $X$. In Fig. \ref{fig:korteTijden}, we plotted this distribution for early times $t \ll \tau$ for $T=0$ and $T=1$. Fig. \ref{fig:langeTijden} shows the same distribution for larger times $\tau_N \ll t < \tau$, again for $T=0$ and $T=1$. For times of the order of $\tau_N$ the distribution is clearly non-Gaussian.  We see that  the distribution is almost uniform in this regime. When the temperature is switched off, this effect can be seen from $t=0$, whereas for non-zero temperature this effect builds up towards $t\approx\tau_N$. When time increases towards the inverse tumble rate the distribution becomes Gaussian as already noticed in the kurtosis. In the late time regime there is little difference between the distributions at $T=0$ and $T=1$. 

The distribution for early times and at $T=0$ can be easily understood as follows. In this regime, most monomers move along a straight line with constant velocity $v$. Since the direction of this motion is random we can expect the $x$-component of the displacement to be uniformly distributed between $-vt$ and $vt$. This ballistic motion becomes modified as soon as the effect of the springs becomes important, i.e. from $t \approx \tau_N$ on. 

In conclusion, we find that the early time regimes are non-Gaussian. At $T=0$ this is the ballistic regime and the superdiffusive one. The ballistic regime is always very short since $\tau_N \sim \gamma/k$, i.e. it does not depend on $N$. In contrast $\tau_R \sim N^2$. The requirement $\tau < \tau_R$ therefore allows the possibility of a non-Gaussian superdiffusive regime that can last for times of order $N^2$. 

\section{Lattice model}
The study of lattice gases of driven particles has a long history in non-equilibrium statistical physics. Among the most studied models are exclusion processes where particles hop on a lattice subject to an exclusion principle, i.e. at each lattice site there can be at most one particle \cite{Spitzer70,MacDonald68,Derrida07,Chou11,Gorissen12}. Exclusion processes including active particles or active particles in a bath of passive ones have been studied recently \cite{Bertrand18,Soto14,Solon13,Manacorda17,Thompson11,Whitelam18,Kourbane18}. They show unexpected phenomena such as cluster formation and mobility induced phase transitions \cite{Fily12,Cates15}, in which collections of active particles with purely repulsive interactions phase separate in regions of high and low density, a phenomenon that can only occur out of equilibrium. Here, we do not want to investigate these collective phenomena but focus on the diffusive properties of a tagged active particle in a bath of other active particles.

\begin{figure}
\centering
\includegraphics[width=8cm]{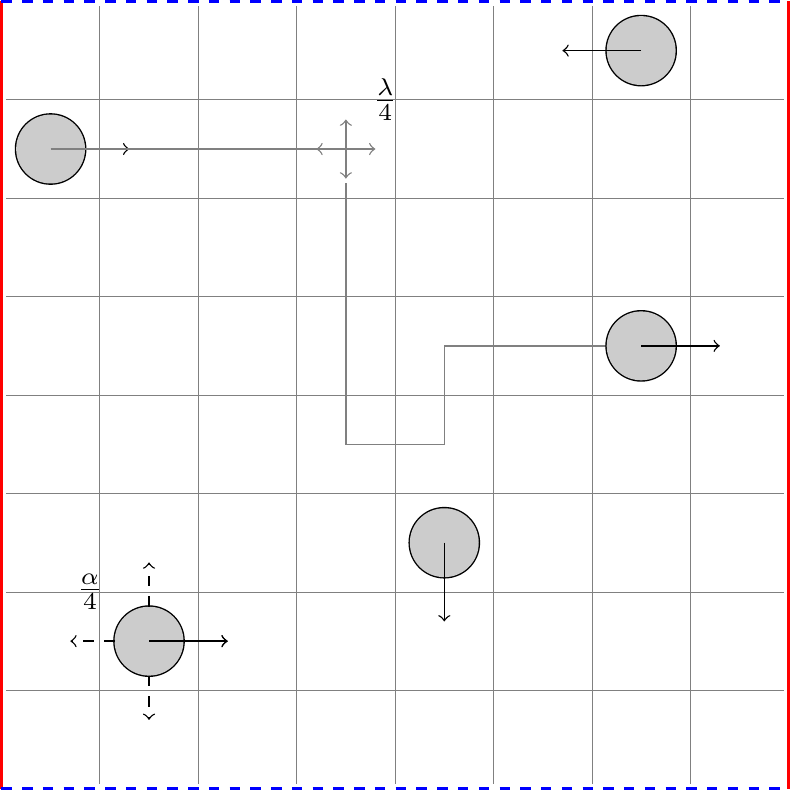}
\caption{Lattice model of active particles. The particles move in a fixed direction (black arrow), which changes with rate $\lambda$ (grey arrows). Besides this the particles can jump to any neighbouring site with rate $\alpha$ while keeping their direction of motion fixed (dashed arrows).}
\label{fig6}
\end{figure}

Our model is a variant of the persistent exclusion process (PEP) which was introduced in \cite{Soto14}. It is defined on an $L \times L$ square lattice (see Fig. \ref{fig6}). We use periodic boundary conditions. On the lattice particles can jump to neighbouring sites provided that they are empty. To model run-and-tumble dynamics at $T=0$, each particle has a fixed direction (velocity) $\vec{e}$ which can assume four values (up, left, down or right ). Only jumps in the direction of $\vec{e}$ are allowed. This motion corresponds with runs. With a (small) rate $\lambda$, the particle tumbles, after which its velocity is chosen again at random from the four allowed values. To model the effect of temperature, the particle can also move to any neighbouring site with rate $\alpha$ while keeping $\vec{e}$ fixed. We are interested in the low temperature regime $\alpha < \lambda$ where the active motion dominates. 

We have performed extensive simulations of this model for $L=40$. We follow the motion of a tagged particle for which we monitor one component, say the $x$-component, of the position. We are interested in the statistical properties of this quantity. Again we look at the variance $\Delta x^2(t)�\equiv \langle (x(t)-x(0))^2�\rangle$, the kurtosis $\kappa(t)\equiv\langle (x(t)-x(0))^4 \rangle/(\langle (x(t)-x(0))^2\rangle)^2$ and the whole probability distribution of $x(t)-x(0)$. Our results are obtained after averaging over 25000 histories. 

We first discuss the results at zero temperature, i.e. for $\alpha=0$. We take $\lambda=0.005$ so that the associated persistence time for the run-and-tumble particles is $\tau=1/\lambda=200$. As could be expected, the results depend crucially on the density of the particles.

\begin{figure}
\centering
  \includegraphics[width=7.5cm]{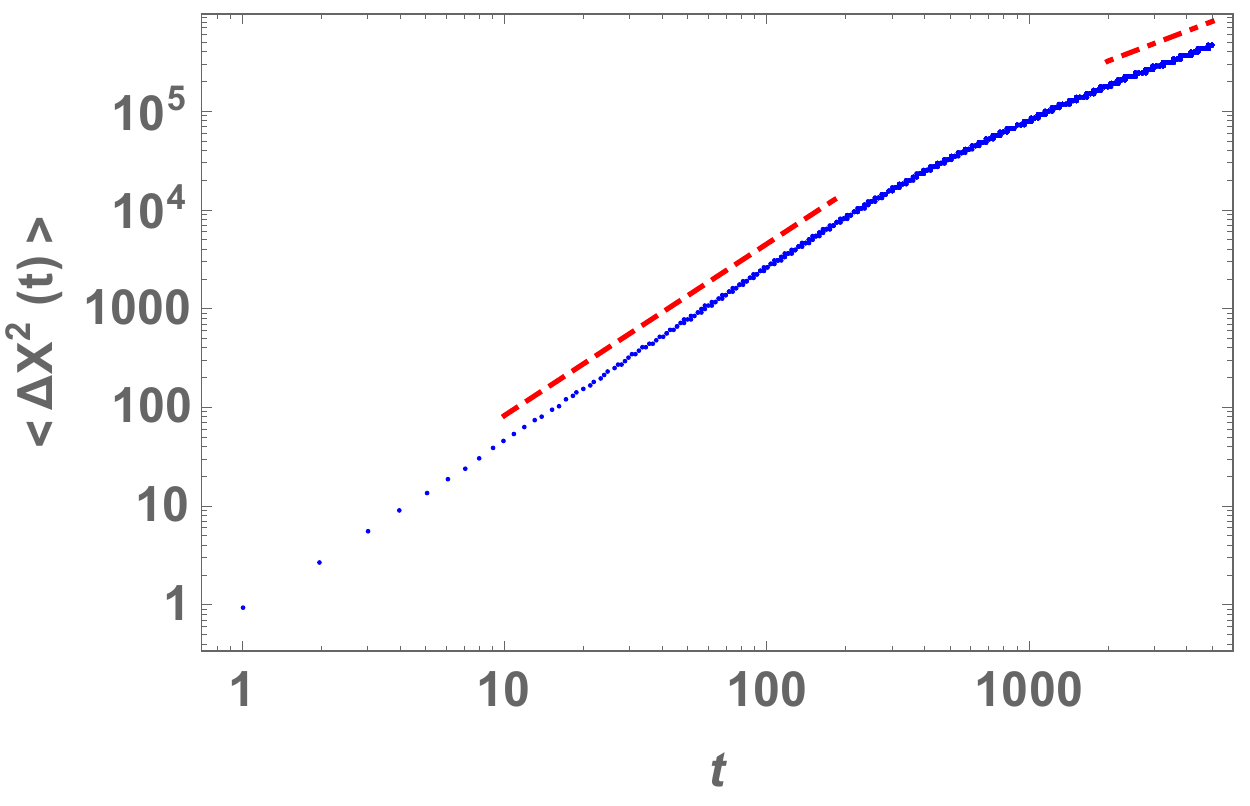}
  \includegraphics[width=7.5cm]{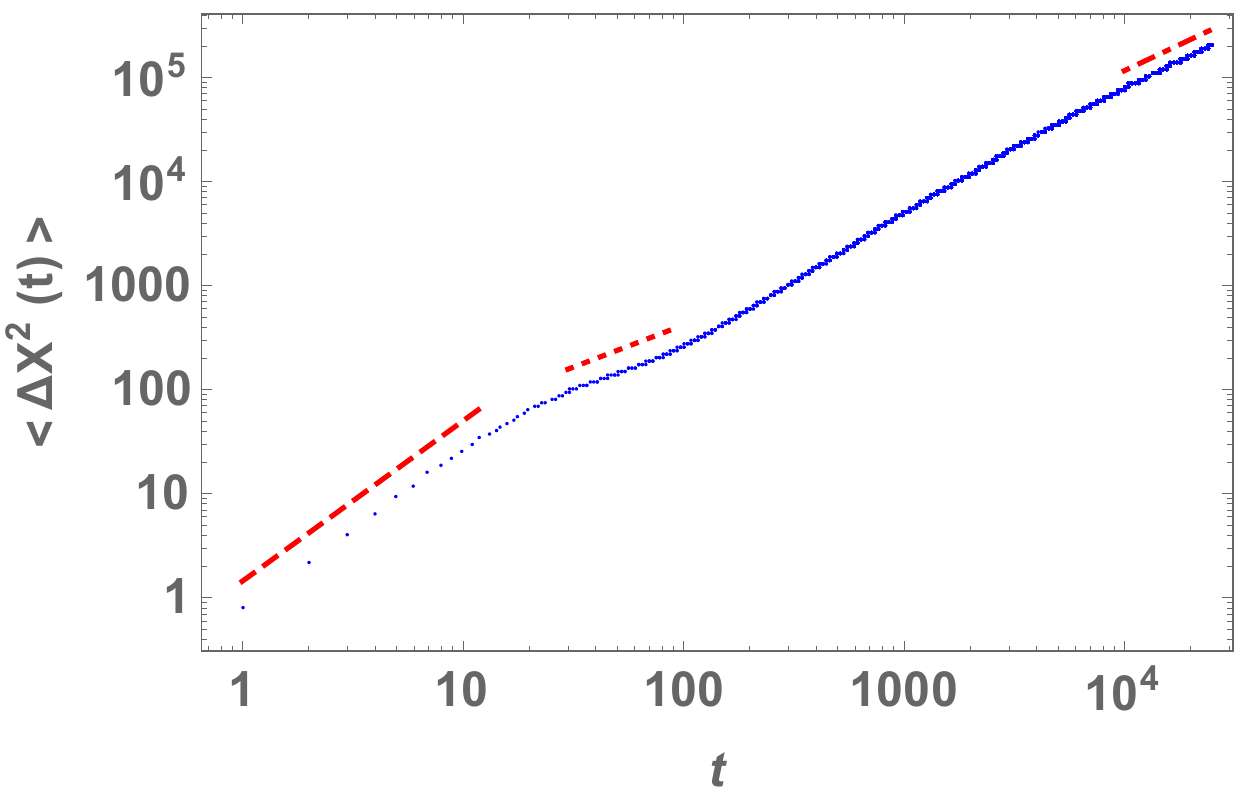}
\caption{Log-log plot of the variance of the $x$-component of the position of the tagged particle at low density $\rho=0.05$ (left side) and intermediate density $\rho=0.25$ (right side). At low density the behaviour crosses over from superdiffusion to normal diffusion. At intermediate density a subdiffusive regime appears. The dashed line has slope $1.74$ (left) and $1.54$ (right) while the dotdashed line has slope $1.01$ (left) and $1.00$ (right). Finally the dotted line has slope $0.81$. }
\label{fig7}
\end{figure}

We show our results for the variance of the position in Fig. \ref{fig7} for a low density of particles, $\rho=0.05$ (left side), and for an intermediate density $\rho=0.25$ (right side)\footnote{At higher density the time evolution of the model slows down considerably which makes it difficult to obtain reliable results for late times.}. As can be seen, at low density there is a superdiffusive regime for $t < \tau$. The exponent of this regime is estimated as $1.74$ (dashed line). Then there is a crossover to a diffusive regime. This crossover takes a very long time and it is only for times larger than $10 \tau$ that the exponent reaches a value $1.01$ (dot-dashed line). If we increase the density, the duration of the superdiffusive regime becomes shorter and the crossover to the diffusive regime takes much longer. The exponent of the superdiffusive regime was found to decrease to $1.54$ at $\rho=0.25$ (dashed line). More interestingly, there appears an intermediate subdiffusive regime. This regime lasts untill $t \sim \tau$.  For $\rho=0.25$ the exponent in this regime equals $0.81$ (dotted line). When we increase the temperature by making $\alpha$ different from zero, we find that there is no qualitative change as long as $\alpha$ is not much larger than $\lambda$. We therefore concentrate in our further discussion on the case $\alpha=0$. 

\begin{figure}
\centering
  \includegraphics[width=7.5cm]{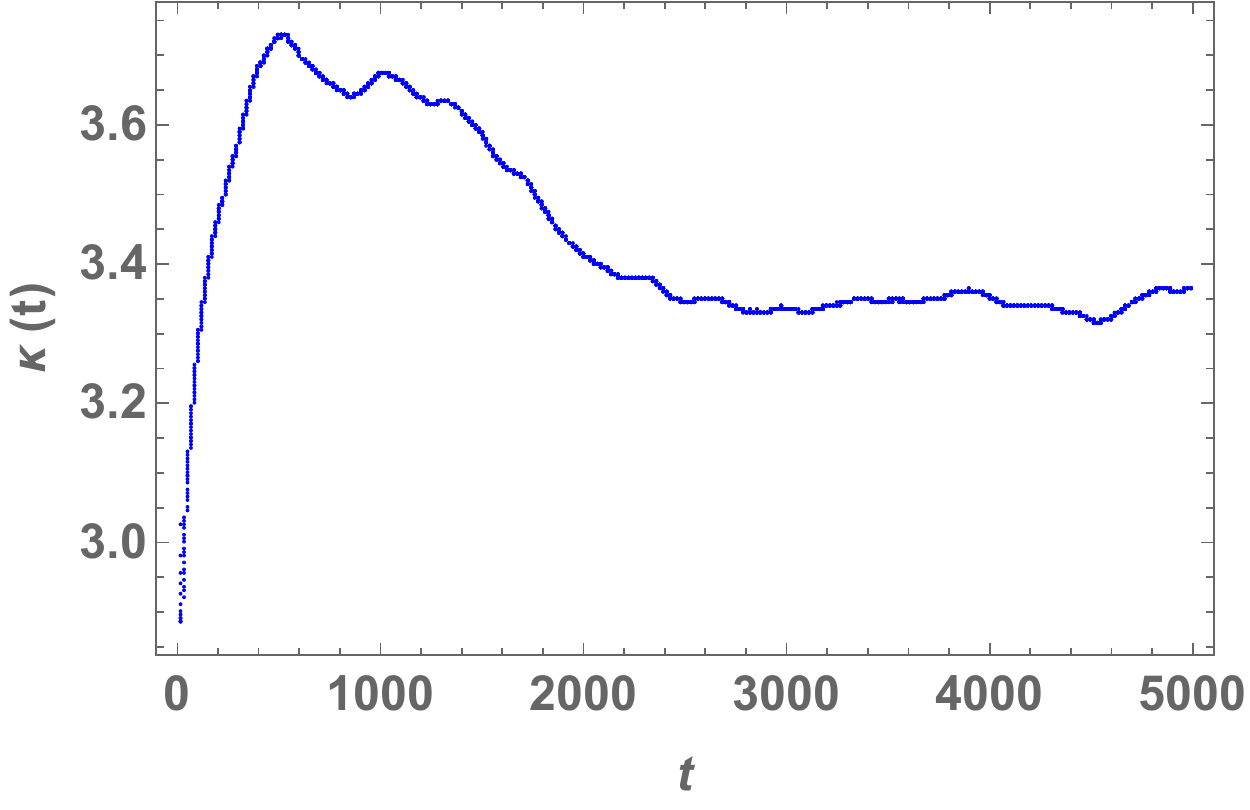}
  \includegraphics[width=7.5cm]{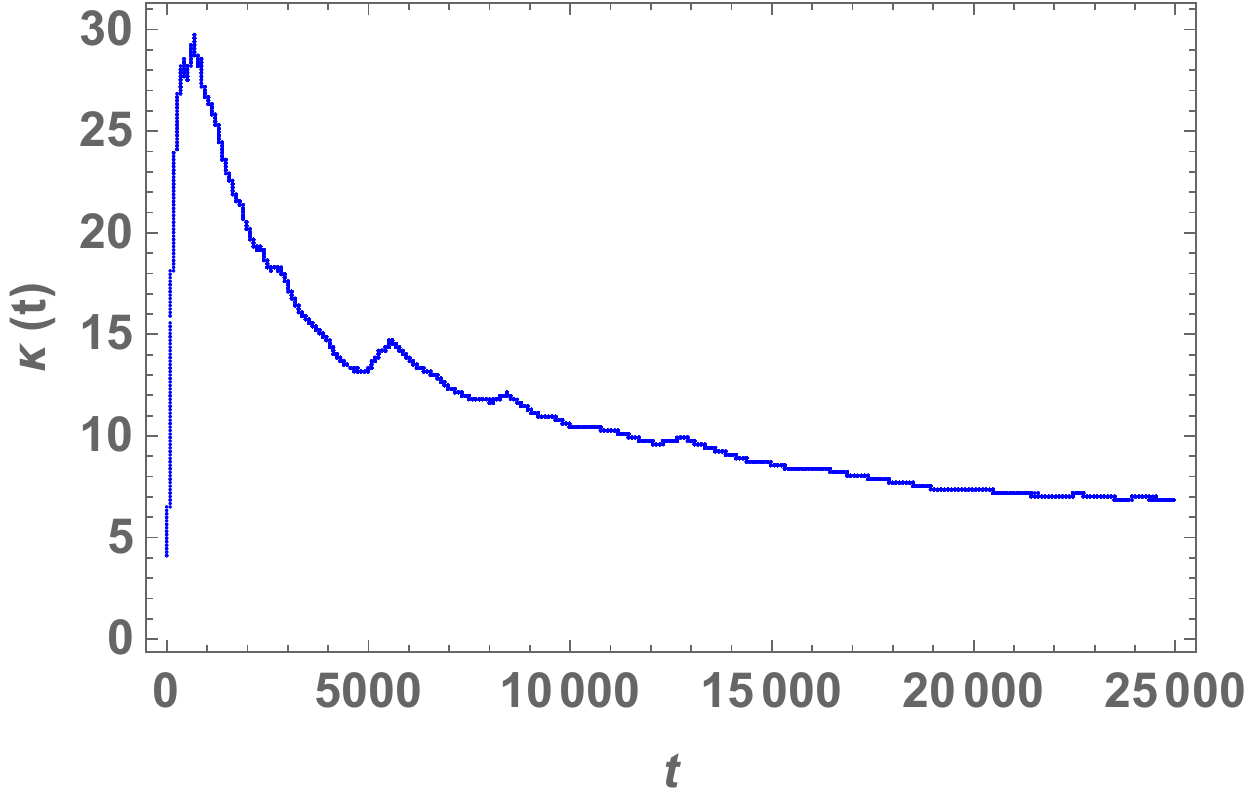}
\caption{Kurtosis $\kappa(t)$ of the $x$-component of the position of the tagged particle at low density $\rho=0.05$ (left side) and intermediate density $\rho=0.25$ (right side). Both plots are made for $T=0$.}
\label{fig8}
\end{figure}

The motion of the tagged particle at intermediate densities is at first sight very similar to that of the tagged monomer. However, important differences appear when we go beyond the second moment and look also at the kurtosis. In Fig. \ref{fig8} we plot this quantity for a low density (left) and an intermediate density (right).
At low-density, the kurtosis increases for times less than the persistence times $\tau$ (i.e. in the superdiffusive phase) after which it decreases and reaches a stationary value in the diffusive regime.  For all times, there is however a clear deviation from Gaussianity. For times $t>10 \tau$ (diffusive regime), the kurtosis has a constant value of approximately $3.3$. This constitutes clear evidence for a diffusive, non-Gaussian behaviour. It is an interesting question whether this regime is the stationary one or whether at much longer time scales a second crossover to a Gaussian diffusive regime occurs. At intermediate density (Fig. \ref{fig8}, right), a similar behaviour is observed. The kurtosis increases in the early time super- and subdiffusive regimes after which it starts a slow decrease.  The deviations from Gaussianity as parametrised by the kurtosis are always much larger than at low density. The kurtosis keeps on decreasing down to $t \approx 125 \tau$ where its value is approximately $6.8$. 

\begin{figure}
\centering
  \includegraphics[width=7.5cm]{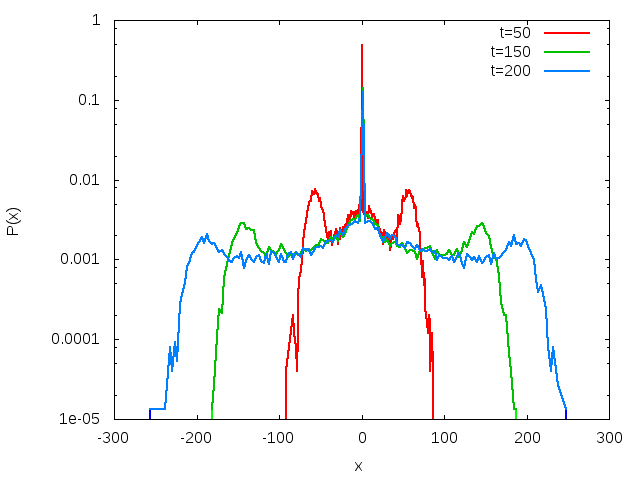}
  \includegraphics[width=7.5cm]{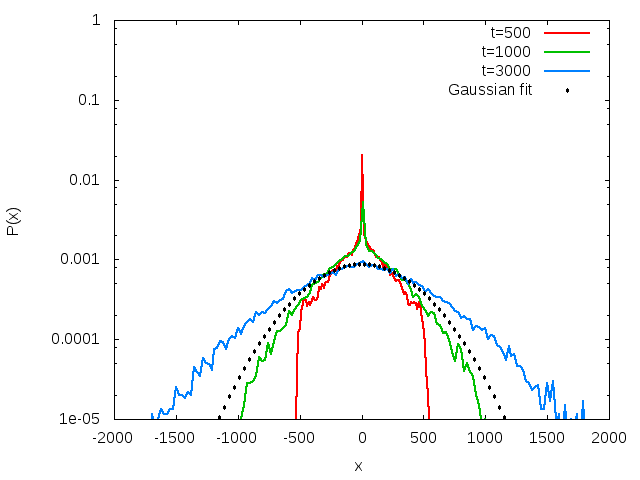}
\caption{Probability distribution $P(X)$ of the $x$-component $X=x(t)-x(0)$ of the position of the tagged particle. The data are for $T=0$ and $\rho=0.05$. The figure on the left shows $P(X)$ at three times that are smaller than the persistence time $\tau=200$, while that on the right plots $P(X)$ for times larger than $\tau$. The dotted line presents a Gaussian fit through the data for $t=3000$.}
\label{fig9}
\end{figure}

We finally look at the probability distribution of the $x$-component of the position of the tagged particle. Fig. \ref{fig9} shows our results for the low density system for times less than the persistence time (left) and for times greater than $\tau$ (right). For all simulated times the behaviour is non-Gaussian. 
For the largest time ($t=200 \tau$) shown we have fitted the distribution with a Gaussian. This fit is good for small values of the random variable but clearly there are non-Gaussian tails, consistent with what we found from the kurtosis. The early time behaviour of this low-density system can be easily understood. For times that are smaller than $\tau$ most of  the particles run in a fixed direction. The $x$-component therefore grows proportional to $t$ ($-t$) for a particle running right (left) while it remains zero for a particle running up or down. We therefore expect three peaks at these early times which is indeed what is observed for $t=50$ and to a lesser extend at $t=150$ or $t=\tau$ when both the tumbling of the particles and the presence of other particles modifies this simple motion. There remains a peak in the distribution at $x=0$ even for times much larger than $\tau$ indicating that there is still a subset of particles that didn't tumble yet. Only for times $t>10\tau$ this peak has disappeared and in this regime the motion becomes diffusive, yet with non-Gaussian tails. 

We now turn to the probability distribution of the position at higher density. For earlier times the distribution is narrower than at low density and the peaks have disappeared. This can be understood as follows: for times less than $\tau$ the particles move on a straight line but due to the higher density the particles cannot move for a large fraction of $t$. The actual distance travelled is strongly dependent on the local environment and hence the distance travelled is distributed between $0$ and $t$. This behaviour is  also reflected in the subdiffusion evident from the variance of the position.
\begin{figure}
\centering
  \includegraphics[width=7.5cm]{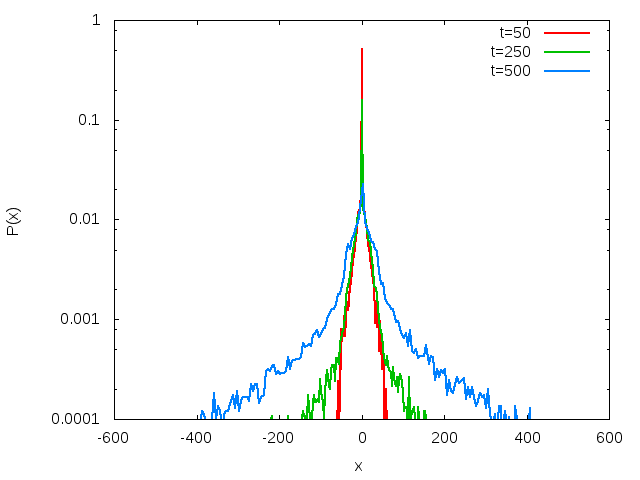}
  \includegraphics[width=7.5cm]{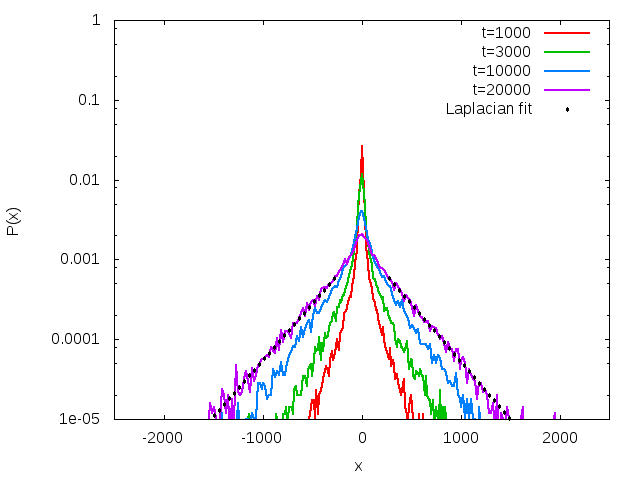}
\caption{Probability distribution $P(X)$ of the $x$-component  $X=x(t)-x(0)$ of the position of the tagged particle. The data are for $T=0$ and $\rho=0.25$.  The figure on the left shows $P(X)$ at three times that are smaller than or of the order of the persistence time $\tau=200$. The figure on the right shows $P(X)$ for times larger than $\tau$. The dotted line presents a fit with a Laplace distribution through the data for $t=20000$.}
\label{fig10}
\end{figure}
In Fig. \ref{fig10}, right side, we see that for large times, $\ln [P(X)] \sim X$ for $X$ not too small (dotted line). A well-known distribution that has this property is the Laplace distribution
\begin{eqnarray}
P(X) = \frac{1}{\sqrt{2} \sigma} \exp\left(-\frac{|X-\mu | \sqrt{2}}{\sigma}\right)
\end{eqnarray}
where $\mu$ and $\sigma$ are the average and the variance of the distribution. The kurtosis of the Laplace distribution equals $6$. 

A Laplace distribution has been observed in various experiments \cite{Wang09,Toyota11,Weeks00,Leptos09}. 
For example, in \cite{Lampo17} it was found that the displacement of RNA-protein particles in {\it E. Coli} is subdiffusive and that its distribution is well approximated by the Laplace distribution. The kurtosis was found to be in the range $4 \sim 5$. Our simulations indicate that for large times also the displacement of a tagged particle in the PEP could be approximated by such a distribution. There are deviations from Laplace behaviour at small displacements, which are also reflected in a somewhat larger value of the kurtosis. 

\begin{figure}
\centering
\includegraphics[width=7cm]{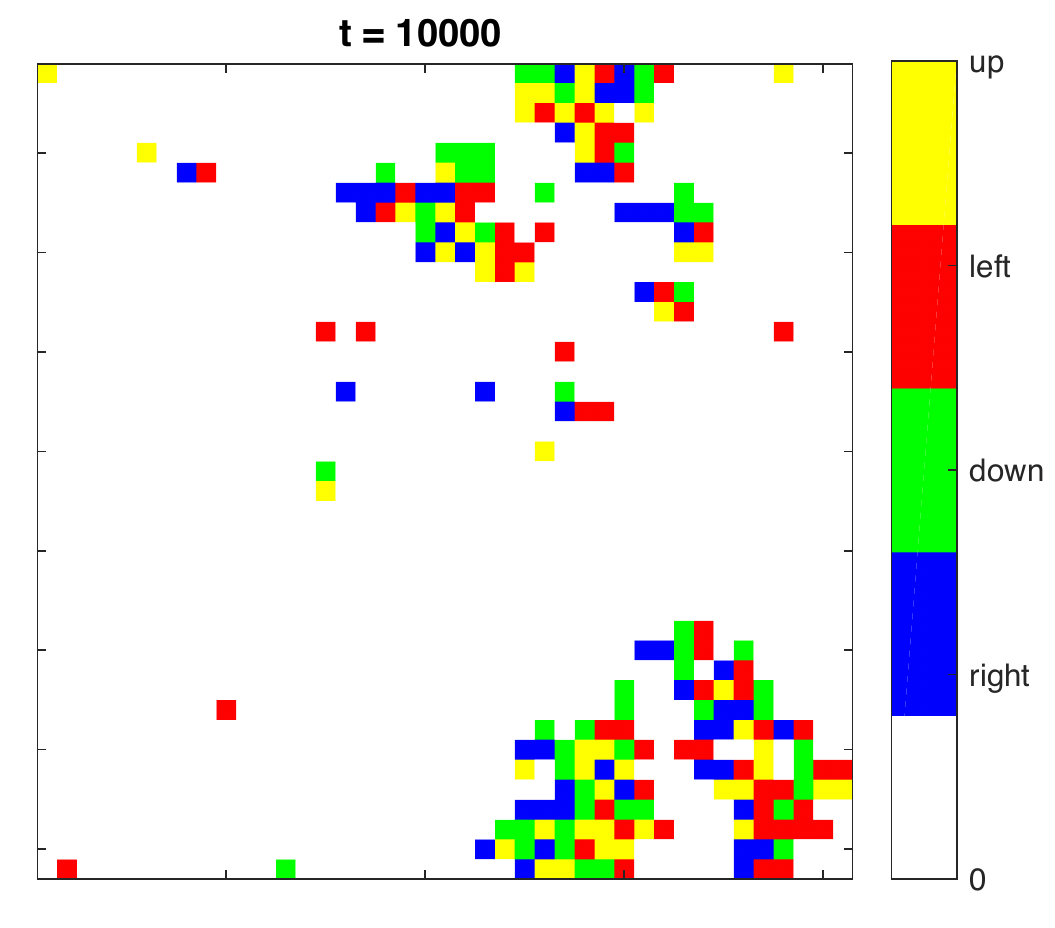}\includegraphics[width=7cm]{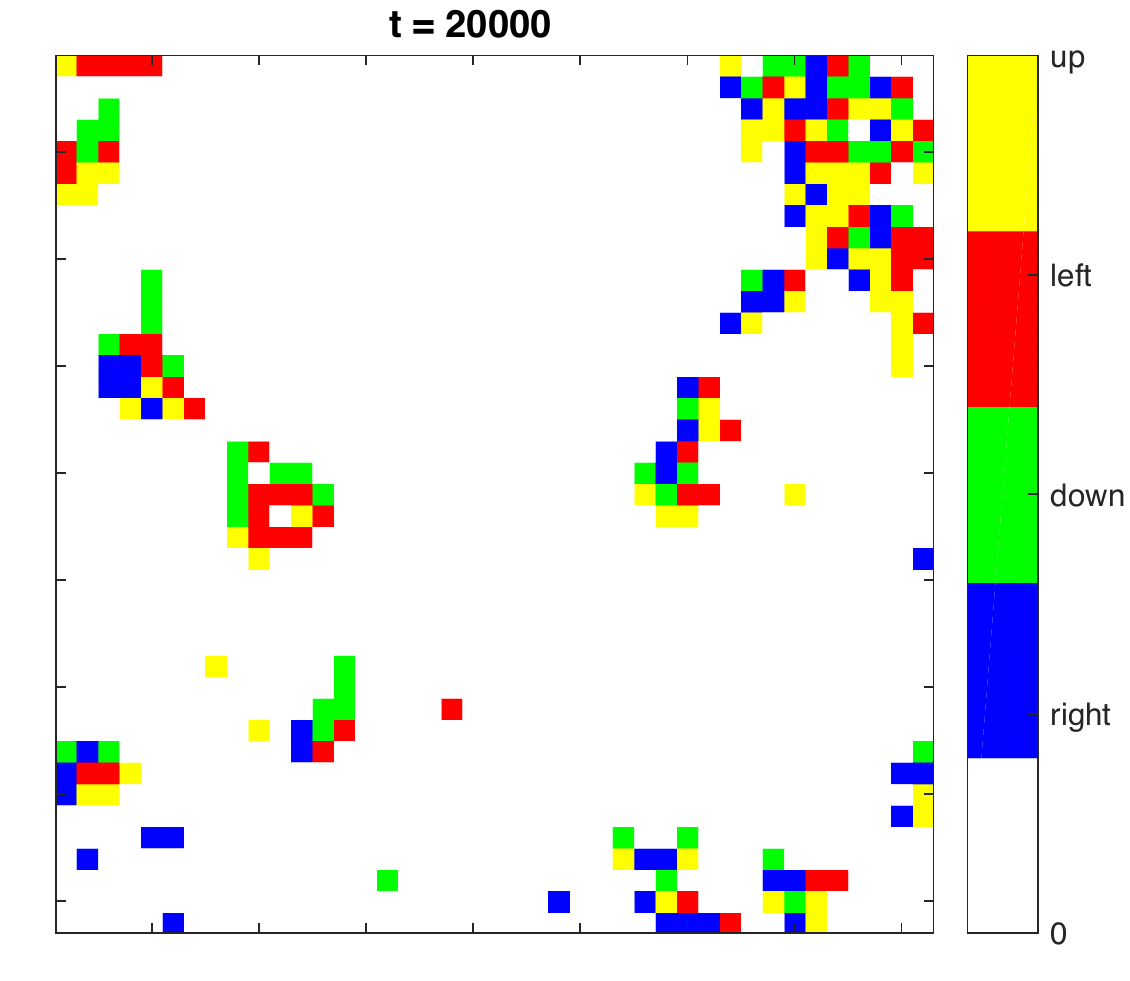}
\caption{Typical configurations of the lattice at density $\rho=0.25$. Empty sites are indicated in white. Occupied sites are coloured. The colour indicates the direction of the velocity of the particle.}
\label{fig11}
\end{figure}

The appearance of a Laplace distribution can be understood in terms of the diffusing diffusivity model \cite{Chubynsky14,Jain16,Chechkin17}. The basic physical idea is that the diffusivity $D$ of a particle diffusing in a complex medium evolves in time due to the heterogeneities of the environment. This is described by a time dependent diffusivity $D(t)$ that also evolves stochastically. For example in the model of \cite{Chechkin17}, $D(t)=Y^2(t)$ where $Y(t)$ is an Ornstein-Uhlenbeck process. In our lattice model, we observe at intermediate densities a continuous appearance and disappearance of clusters \cite{Soto14}. For example, in Fig. \ref{fig11} we show two typical configurations of the lattice. Different colours indicate different velocities. It is clear from this figure that most of the particles in a cluster cannot move because there is another particle in the direction of their velocity. A particle can then only move when has moved to the surface of a cluster and then tumbles. Numerical studies of the PEP \cite{Soto14} has shown that the distribution of cluster sizes is the product of a power law and exponential. Thus an active particle may at some time move freely while at another time its motion is temporarily stopped when it is in a cluster. The time it remains stopped in a cluster has a broad distribution \cite{Soto14}. In this way, the dynamics of the model can create a heterogeneous environment for the particle which could explain, at least heuristically, the observed distribution of its displacement. 

\section{Conclusions}
In this paper, we have studied two models of interacting run-and-tumble particles. In the first one, the particles form a polymer and interact with a harmonic force. In the second one, the particles move on a lattice and interact through exclusion. We were interested in the motion of a tagged particle for which we determined the statistical properties of its displacement.

Both models showed regimes of superdiffusive, subdiffusive and diffusive motion. The main difference is that in the polymer model only the early time regime (lasting up to the persistence time of the run-and-tumble motion) was found to be non-Gaussian. In the lattice model, this non-Gaussianity persisted for all time scales simulated. The non-Gaussianity observed at times below the persistence time in both models can be easily understood in terms of the ballistic motion that the particles perform for times less than $\tau$. More interesting is the time regime $t > \tau$ for the lattice model at intermediate density where evidence for a Laplace distribution of the displacement was found. 

The polymer model is probably too simple to describe true non-equilibrium behaviour for $t > \tau$. On time scales larger than the persistence time $\tau$, the active forces $\vec{\eta}_n(t)$ become delta-correlated and their effect can therefore be added to those of the thermal forces. In this way, the run-and-tumble polymer becomes equivalent to an equilibrium Rouse model albeit at an effective, higher, temperature. This can explain the observed Gaussianity. This is in contrast with polymer models where active forces act heterogeneously along the chain \cite{Gladrow16,Put19}. On time scales above the persistence time, such a model can be mapped on a Rouse chain with a non-homogeneous temperature which clearly is a system that is out of equilibrium. However even in such a model preliminary simulations show Gaussian behaviour at late times. Most probably this is because the environment that a tagged monomer sees remains the same due to the chain structure of the polymer. For this reason we also do not expect non-Gaussianity in the dynamics of other reaction coordinates  (end-to-end distance, number of bounded monomers,...)  for polymers out of equilibrium. In contrast,  in the lattice model, especially at intermediate densities, the environment is fluctuating. As soon as the typical time between particle interactions becomes smaller than the persistence time, the motion of a tagged particle has periods of free ballistic motion that alternate with periods in which the motion is slowed down or stopped because the particle is in a cluster. Therefore it seems that a dynamically changing environment is essential to observe  non-Gaussian behaviour in the motion of a slow coordinate in active systems. 

\section{Acknowledgements}
We wish to acknowledge the work and effort of Carlo Vanderzande that he put into this and many other research projects. He passed away on September 2, 2019. We will always remember his kindness, his never ending enthusiasm for research and his extensive knowledge. We grieve the loss of a great friend and colleague.
\clearpage
\newpage

{\bf References}\\

\newpage
\appendix

\setcounter{equation}{0}
\setcounter{figure}{0}
\setcounter{table}{0}
\makeatletter

\section{Time evolution and correlation of the normal modes}\label{appendix:normModes}

To determine the time evolution of the normal modes, we need to solve the Langevin equation (\ref{eq:Langevin}). In order to do this, we take the time-derivative of the normal coordinates and multiply it by $\gamma$:

\begin{equation}
    \gamma \frac{d}{dt} \vec{X}_p(t) = \frac{\gamma}{N} \sum_{n=0}^{N-1} c^p_n \frac{d}{dt}\vec{R}_n(t).
\end{equation}
Using equation (\ref{eq:Langevin}), we can rewrite this expression:
\begin{eqnarray}
    \gamma \frac{d}{dt} \vec{X}_p(t) & & = -\frac{k}{N}\sum_{n=0}^{N-1}c^p_n\left[2 \vec{R}_n(t) - \vec{R}_{n-1}(t) - \vec{R}_{n+1}(t) \right] \nonumber  \\ & &+ v \sum_{n=0}^{N-1}c^p_n \hat{e}_n(t) + \sum_{n=0}^{N-1}c^p_n\vec{\eta}_n(t).
\end{eqnarray}
Next, we replace the natural coordinates in the above expression with formula (\ref{eq:naturalcoords}). After rearranging some terms, we get the following equation:
\begin{eqnarray}
    \gamma \frac{d}{dt} \vec{X}_p(t) & & = -\frac{8k}{N}\sum_{q=1}^{N-1}\sin^2\bigg(\frac{\pi q}{2N}\bigg)\vec{X}_q(t)\sum_{n=0}^{N-1}c^p_nc^q_n \nonumber  \\ & &+ v \sum_{n=0}^{N-1}c^p_n \hat{e}_n(t) + \sum_{n=0}^{N-1}c^p_n\vec{\eta}_n(t).
\end{eqnarray}
By using 
\begin{equation}\label{eq:orthoCoef}
\sum_{n=0}^{N-1}C^p_nC^q_n = \frac{N}{2}\left(1+\delta_{p,0}\right)\delta_{p,q}
\label{a4}
\end{equation}
and after replacing the last two terms by $v\vec{e}_p(t)$ and $\vec{\eta}_{p}(t)$, we get a new Langevin equation for the normal coordinates:
\begin{equation}
    \gamma \frac{d}{dt} \vec{X}_p(t) = -\frac{k_p}{2N}\vec{X}_p(t) + v\vec{e}_p(t) + \vec{\eta}_{p}(t),
\end{equation}
with $k_p=8 N k \sin^2 \left(\pi p/2 N\right)$.
After solving this differential equation, we get the time evolution for the normal mode $\vec{X}_p(t)$:
\begin{equation}
    \vec{X}_p(t) = \vec{X}_p(0)e^{-t/\tau_p} + \frac{v}{\gamma}\int_0^t\vec{e}_p(t-\tau)e^{-t/\tau_p}d\tau + \frac{1}{\gamma}\int_0^t\vec{\eta}_p(t-\tau)e^{-t/\tau_p}d\tau.
\end{equation}

The normal thermal force $\vec{\eta}_{p}(t)$ is again a Gaussian random variable with zero average and a correlation which can be obtained from (\ref{eq:thermal}): 
\begin{equation}\label{eq:thermcor}
\langle \vec{\eta}_{p}(t)\cdot\vec{\eta}_{q}(t') \rangle = \frac{3k_BT}{N} \gamma (1+\delta_{p,0})\delta_{p,q} \delta(t-t'),
\end{equation}
where we have used the orthogonality relation (\ref{a4}).
On the other hand, the normal active forces have zero average and correlation 
\begin{eqnarray}
\langle \vec{e}_p(t)\cdot \vec{e}_q(t') \rangle & & = \frac{v^2}{N^2}\sum_{n = 0}^{N-1}\sum_{m = 0}^{N-1}c^p_n c^q_m \langle \vec{e}_n(t)\cdot \vec{e}_m(t') \rangle \nonumber \\ & & = \frac{v^2}{2N} e^{-\lambda |t'-t|}\delta_{p,q}.
\label{eq:activcor}
\end{eqnarray}
At this moment, we have all ingredients to calculate the correlation of the normal modes. 
\begin{eqnarray}\label{eq:corRouseModes}
\langle\vec{X}_p(t)\cdot\vec{X}_q(t')\rangle & & = \langle\vec{X}_p(0)\cdot\vec{X}_q(0)\rangle e^{-t/\tau_p} e^{-t'/\tau_q} \nonumber \\ & & + \frac{1}{\gamma^2}\int_0^td\tau \int_0^{t'}\Big[v^2\langle\vec{e}_{p}(t-\tau)\cdot\vec{e}_{q}(t'-\tau')\rangle \nonumber \\ & & + \langle\vec{\eta}_{p}(t-\tau)\cdot\vec{\eta}_{q}(t'-\tau')\rangle\Big] e^{-\tau/\tau_p} e^{-\tau'/\tau_q}d\tau' ,
\end{eqnarray}
where we used the fact that the initial value of the normal coordinates, the normal thermal noise and the normal active noise are uncorrelated. 
Filling in the equipartition property for the initial value of the normal coordinates, and the expressions for the correlations described above, we eventually get the following formula for the correlation of the normal modes:
\begin{eqnarray}\label{eq:XpXq}
\langle\vec{X}_p(t)\cdot\vec{X}_q(t')\rangle = \frac{3k_B T}{k_p} e^{- |t-t'|/\tau_p}\ \delta_{p,q} + \frac{v^2}{2N\gamma^2}
C_{p}(t,t')\delta_{p,q},
\end{eqnarray}
with
\begin{eqnarray}
C_{p}(t,t') = \int_0^t d\tau \int_0^{t'} e^{-\lambda|t-\tau-t'+\tau'|} e^{-(\tau+\tau')/\tau_p}d\tau'.
\label{integrals}
\end{eqnarray}

\section{Mean squared displacement}\label{appendix:msd}
To obtain the correlation function for the monomer position, one can use the formula for the position vector as function of time for the $n$th monomer:
\begin{eqnarray}
\Delta \vec{R}_n(t) & & \equiv \langle \vec{R}_n(t) - \vec{R}_n(0) \rangle \nonumber \\ 
& & = \vec{X}_0(t) + 2\sum_{p=1}^{N-1}C^p_n\left[\vec{X}_p(t)-\vec{X}_p(0)\right]. \label{eq:positionR}
\end{eqnarray}
Taking the square gives us the expression for the mean squared displacement:
\begin{eqnarray}
\Delta R_n^2(t) & & = \langle [\vec{R}_n(t) - \vec{R}_n(0) ]^2\rangle \nonumber  \\ & & = \langle \vec{X}_0^2(t) \rangle \nonumber \\ & & + 4\sum_{p=1}^{N-1}\sum_{q=1}^{N-1}c^p_nc^q_n\langle[\vec{X}_p(t)-\vec{X}_p(0)]\cdot[\vec{X}_q(t)-\vec{X}_q(0)]\rangle \nonumber \\ & & + 2\sum_{p=1}^{N-1}c^p_n\langle[\vec{X}_p(t)-\vec{X}_p(0)]\cdot\vec{X}_0(t)\rangle \nonumber \\ & & + 2\sum_{q=1}^{N-1}c^q_n\langle[\vec{X}_q(t)-\vec{X}_q(0)]\cdot\vec{X}_0(t)\rangle.
\end{eqnarray}
Picking $t=t'$, we choose the initial condition $\vec{X}_0(0)=0$ and by using the equations for the correlation between thermal/active normal modes and the Rouse modes (equations (\ref{eq:thermcor}), (\ref{eq:activcor}), (\ref{eq:XpXq})), one obtains the following contributions to the mean squared displacement:
\begin{eqnarray}
\langle \vec{X}_0^2(t) \rangle = \frac{6k_BT}{N\gamma}t + \frac{v^2}{N\gamma^2}I_0,
\end{eqnarray}
\begin{eqnarray}
4\sum_{p=1}^{N-1}\sum_{q=1}^{N-1}c^p_nc^q_n\langle\vec{X}_p(t)\vec{X}_q(t)\rangle = 4\sum_{p=1}^{N-1}(c^p_n)^2\Bigg[\frac{3k_BT}{k_p} + \frac{v^2}{2N\gamma^2}C_p(t,t)\Bigg],
\end{eqnarray}
\begin{eqnarray}
4\sum_{p=1}^{N-1}\sum_{q=1}^{N-1}c^p_nc^q_n\langle\vec{X}_p(0)\vec{X}_q(0)\rangle = 4\sum_{p=1}^{N-1}(c^p_n)^2\frac{3k_BT}{k_p},
\end{eqnarray}
\begin{eqnarray}
4\sum_{p=1}^{N-1}\sum_{q=1}^{N-1}c^p_nc^q_n\langle\vec{X}_p(t)\vec{X}_q(0)\rangle = 4\sum_{p=1}^{N-1}(c^p_n)^2\frac{3k_BT}{k_p}e^{-t/\tau_p}.
\end{eqnarray}
Here 
\begin{eqnarray}\label{eq:I0}
I_0(t)=\int_0^t d\tau \int_0^{t} e^{-\lambda |\tau-\tau'|} d\tau'=\frac{2(-1+\lambda t + e^{-\lambda t})}{\lambda^2}.
\end{eqnarray}
All other mixed terms are zero due to formula (\ref{eq:orthoCoef}) and the fact that $p$ and $q$ start from the value 1. All together, these contributions lead to the final expression for the mean squared displacement:
\begin{eqnarray}
\Delta R_n^2(t)  &=& \frac{6k_B T}{N \gamma}\left[t + 2 \sum_{p=1}^{N-1} \left(c_n^p\right)^2 \tau_p \left(1-e^{-t/{\tau_p}}\right)\right] \nonumber \\
&+& \frac{v^2}{N \gamma^2}\left[I_0(t) + 2 \sum_{p=1}^{N-1} \left(c_n^p\right)^2  C_p(t,t)\right].
\label{eq:msd}
\end{eqnarray}
The first two terms are the thermal contribution to the correlation, the last two terms represent the active part. As a matter of fact, the first and third terms represent center of mass movement.

\section{Superdiffusive behaviour for $\tau_N < t < \tau$}\label{appendix:superdiff}

Starting with the active contribution to equation (\ref{eq:msd}) for the mean squared displacement, we can see that for $t < \tau$, the contribution (\ref{eq:I0}) for $I_0(t)$ becomes zero. Furthermore, we have
\begin{eqnarray}
C_p(t,t) & & \approx \int_0^td\tau\int_0^t e^{-(\tau+\tau')/\tau_p}d\tau' \nonumber \\ & & = \tau_p^2(1-e^{-t/\tau_p})^2.
\end{eqnarray}
If $N$ is large enough, we can approximate the sum in the active part of equation (\ref{eq:msd}) by an integral over $p$: 
\begin{eqnarray}
    \int_{0}^{\infty} (C^p_n)^2\tau_p^2(1-e^{-t/\tau_p})^2 dp.
\end{eqnarray}
Using the fact that $\tau_p \approx \frac{a}{p^2}$ (with $a$ independent of $p$) and $C^p_n \approx 1$ for large values of $N$, this integral becomes proportional to
\begin{eqnarray}
    \int_{0}^{\infty} \frac{(1-e^{-tp^2/a})^2}{p^2} dp \sim t^{3/2},
\end{eqnarray}
where we used the variable transformation $tp^2/a=y$ in the last step.

\end{document}